\documentclass{PoS}
\pdfoutput=1

\usepackage{amsmath}
\usepackage{amssymb}
\usepackage{bbm}

\newcommand{\be}{\begin{equation}}
\newcommand{\ee}{\end{equation}}

\DeclareMathOperator{\re}{Re}

\DeclareMathOperator{\spn}{span}
\DeclareMathOperator{\sign}{sgn}
\DeclareMathOperator{\diag}{diag}

\renewcommand{\phi}{\varphi}

\newcommand{\Do}{D_\text{ov}}
\newcommand{\Dw}{D_\text{w}}

\title{Comparing iterative methods to compute the overlap Dirac
  operator at nonzero chemical potential }

\ShortTitle{Comparing iterative methods to compute the overlap Dirac
  operator at $\mu \neq 0$ }

\author{\speaker{Jacques Bloch}, Tobias Breu, and Tilo Wettig\\
        Institute for Theoretical Physics, University of Regensburg, 93040 Regensburg, Germany\\
        E-mail: \email{jacques.bloch@physik.uni-regensburg.de}}

      \abstract{The overlap Dirac operator at nonzero quark chemical
        potential involves the computation of the sign function of a
        non-Hermitian matrix. In this talk we present iterative Krylov
        subspace approximations, with deflation of critical
        eigenvalues, which we developed to compute the operator on
        large lattices. We compare the accuracy and efficiency of two
        alternative approximations based on the Arnoldi and on the
        two-sided Lanczos method. The short recurrences used in the
        latter method make it faster and more effective for realistic
        lattice simulations.}

\FullConference{The XXVI International Symposium on Lattice Field Theory \\
		 July 14 - 19, 2008\\
		 Williamsburg, Virginia, USA}

\begin{document}

\section{The overlap Dirac operator at nonzero chemical potential}

Although chiral symmetry can not be implemented exactly in a
space-time discretization of QCD \cite{Nielsen:1981hk}, a lattice
version of chiral symmetry can be obtained by a renormalization group
blocking transformation. The ensuing lattice chiral symmetry is
embodied by the Ginsparg-Wilson relation \cite{Ginsparg:1981bj}, which
is solved by the overlap Dirac operator proposed by Narayanan and
Neuberger \cite{Narayanan:1994gw,Neuberger:1997fp}.

Astrophysical objects like neutron stars exhibit an abundance of
quarks over anti-quarks, and the study of QCD in such a background
necessitates the introduction of a quark chemical potential $\mu$ in
the QCD Lagrangian. In this situation chiral symmetry still holds and
in Ref.~\cite{Bloch:2006cd} two of the present authors proposed an
extension of the overlap Dirac operator to nonzero chemical potential
which still satisfies the Ginsparg-Wilson relation, 
\be
D_{\text{ov}}(\mu) = \mathbbm{1} + \gamma_5\sign(\gamma_5
D_\text{w}(\mu))\:,
\label{Dovmu} 
\ee 
where $\gamma_5=\gamma_1\gamma_2\gamma_3\gamma_4$ with
$\gamma_1,\ldots,\gamma_4$ the Dirac gamma matrices in Euclidean
space, $\sign$ is the matrix sign function, and
\begin{align}
D_\text{w}(\mu)
 = \mathbbm{1} - \kappa \sum_{i=1}^3 ( T_i^+ + T_i^-) 
- \kappa ( {e^\mu} T_4^+ + {e^{-\mu}} T_4^-)
\end{align}
is the Wilson Dirac operator at nonzero chemical potential
\cite{Hasenfratz:1983ba} with $(T^{\pm}_\nu)_{yx} = (\mathbbm{1} \pm
\gamma_\nu) U_{x,\pm\nu} \delta_{y,x\pm\hat\nu}$, $\kappa =
1/(8+2m_\text{w})$, $m_\text{w} \in (-2,0)$ and $U_{x,\pm\nu}\in$
SU(3).  The exponential factors $e^{\pm\mu}$ implement the quark
chemical potential on the lattice.

For $\mu \ne 0$ the argument $\gamma_5 D_\text{w}(\mu)$ of the sign
function becomes non-Hermitian, and one is faced with the problem of
defining and computing the sign function of a non-Hermitian matrix.
If the matrix $A$ of dimension $N$ is diagonalizable, then $A=U
\diag\{\lambda_i\} U^{-1}$ with eigenvalues $\lambda_i$ and
eigenvector matrix $U$, and a function $f(A)$ can be computed using
the spectral definition \cite{Golub} 
\be f(A) = U
\diag\{f(\lambda_i)\} U^{-1} \:.
\label{fA}
\ee 
If $A$ is not diagonalizable one can still apply an extension of the
spectral definition using the Jordan canonical form.
As the eigenvalues of a general matrix $A$ can be complex, the sign
function computed using Eq.~\eqref{fA} requires the sign of a complex
number, which can be defined by 
\be
\sign(z) \equiv \frac{z}{\sqrt{z^2}} = \sign(\re z) \:,
\ee
where the branch cut of the square root is chosen along the negative
real axis. This definition ensures that $(\sign z)^2=1$ and gives the
usual result when $z$ is real. It is straightforward to check that
this definition ensures that $(\sign A)^2 = \mathbbm{1}$. Therefore the
overlap Dirac operator at $\mu\neq 0$ satisfies the Ginsparg-Wilson
relation, and the operator can have exact zero modes with definite
chirality. The main properties of the operator were discussed in Ref.\
\cite{Bloch:2007xi}.

Since its introduction the operator has been validated in a number of
studies: Its definition is consistent with that of domain wall
fermions at $\mu \neq 0$ when the extension of the fifth dimension is
taken to infinity and its lattice spacing taken to zero
\cite{Bloch:2007xi}, its microscopic density and first peak computed
from quenched lattice simulations on a $4^4$ lattice agree with the
predictions of non-Hermitian chiral random matrix theory
\cite{Bloch:2006cd,Akemann:2007yj}, and the free fermion energy
density was shown to have the correct continuum limit
\cite{Gattringer:2007uu,Banerjee:2008ii}.

\section{Arnoldi approximation for a function of a non-Hermitian matrix}
\label{Sec:Arnoldi}

The exact computation of $\sign(A)$ using the spectral definition
\eqref{fA} is only feasible for small lattice volumes as the memory
requirements to store the full matrix and the computation time needed
to perform a full diagonalization become prohibitively large for
realistic lattice volumes.  Therefore it is necessary to develop
iterative Krylov subspace methods to compute $f(A)b$ for vectors $b$.
For Hermitian matrices the corresponding methods have already reached
a high level of sophistication and are widely applied, but for
non-Hermitian matrices such methods are novel, except for the special
cases of the inverse and the exponential function. The Arnoldi
approximation discussed below was first introduced in Ref.\
\cite{Bloch:2007aw}, while the two-sided Lanczos approximation of
Sec.~\ref{Sec:2sL} was presented in Ref.~\cite{Breu2007}.

Such iterative methods are based on the observation that the unique
polynomial $P_{K}(z)$ of order $K \le N-1$  which interpolates $f(z)$
at all the eigenvalues of $A$ satisfies the equality 
\be
P_{K}(A) b = f(A) b  \quad \text{ for {any} vector $b$} \:,
\ee
as follows from the definition \eqref{fA}.  Therefore it is an obvious
endeavor to construct a good low-degree polynomial approximation for
$y = f(A) b$, taking into account the spectrum of $A$ and the
decomposition of $b$ in the eigenvectors of $A$.

In order to construct such a low order polynomial approximation to
$f(A) b$ we consider the Krylov subspace
$K_k(A,b)=\spn(b,Ab,A^2b,\ldots,A^{k-1}b)$. By definition this
subspace contains all the vectors resulting from the action of an
arbitrary polynomial of degree $\leq k-1$ in $A$ on the source vector
$b$. One of these vectors, namely the orthogonal projection of $f(A)b$
on the Krylov subspace, will minimize $||P_{k-1}(A)b-f(A)b||$ over all
polynomials of degree $\leq k-1$.
  
The Arnoldi method uses the recursive scheme 
\begin{align}
	A V_k = V_k H_k + \beta_k v_{k+1} e_k^T
\end{align}
with 
\begin{align}
V_k^\dagger AV_k=H_k
\label{VAV}
\end{align}
to build an orthonormal basis $V_k=(v_1,\ldots,v_k)$ in $K_k(A,b)$,
where $v_1=b/|b|$, $\beta_k$ is the normalization of $v_{k+1}$, and
$e_k$ is the $k$-th basis vector in $\mathbb{C}^{k}$. The matrix $H_k$
is a Hessenberg matrix (upper triangular + one subdiagonal) of
dimension $k$, whose eigenvalues are called the Ritz values of $A$
w.r.t. $K_k(A,b)$.

Once the Arnoldi basis has been constructed, the projection of 
 $y=f(A)b$ on $K_k(A,b)$ can be written as
\begin{align}
y_\text{proj}= V_k V_k^\dagger f(A) b
	= V_k  V_k^\dagger f(A) V_k V_k^\dagger b \:.
\end{align}
This formal expression requires the knowledge of the exact solution and is therefore of no practical use. However, using the Ritz approximation
\begin{align}
V_k^\dagger f(A) V_k \approx f(H_k) \:,
\end{align}
which is based on Eq.~\eqref{VAV}, allows us to approximate the projection by
\begin{align}
y_\text{proj} \approx \tilde y = |b| V_k f(H_k) e_1 \:.
\label{yprojapp}
\end{align}
By construction this approximation is an element of the Krylov
subspace $K_k(A,b)$, and it can be shown that the implicit polynomial
constructed by this approximation interpolates $f(x)$ at the Ritz
values of $A$ w.r.t. $K_k(A,b)$ \cite{saad:209}. Note that only the
first column of $f(H_k)$ is needed in Eq.~\eqref{yprojapp}.

This approximation reduces the problem to the computation of $f(H_k)$
with $\dim H_k \ll \dim A$, which makes it very useful for practical
use.  The inner sign function $f(H_k)$ will then be computed with
one's method of choice. This could be an exact spectral decomposition
if $k$ is small, or some suitable approximation method, e.g., Roberts'
matrix-iterative method for the sign function,
\[
S^{n+1} = \frac{1}{2} \left[S^n + (S^n)^{-1}\right] \qquad \text{with}
\quad S^0 = H_k \:, 
\] 
which converges quadratically to $\sign(H_k)$.

\section{Sign function and deflation}
\label{Sec:deflation}

When computing the sign function of the matrix $A$ an additional
problem occurs when the matrix has small eigenvalues, as large Krylov
subspaces are then required to achieve a good accuracy. The reason is
that it is not possible to approximate $f$ well by a low-degree
polynomial over the entire spectrum of $A$ if it varies too rapidly in
a small subregion.  A solution to this problem is to use the exact
value of $f$ for a number of critical eigenvalues of $A$. Over the
remaining part of the spectrum $f$ should then behave well enough to
allow for a low-degree polynomial approximation.

Implementing this so-called deflation is straightforward in the
Hermitian case, where it is based on the fact that any number of
eigenvectors span a subspace orthogonal to the remaining eigenvectors.
In the non-Hermitian case the eigenvectors of $A$ are no longer
orthogonal, and a more involved approach is needed.  We have
previously proposed two alternative deflation variants for this case
\cite{Bloch:2007aw}.  Herein we only present the left-right (LR)
deflation and refer to Ref.~\cite{Bloch:2007aw} for the details of
the Schur deflation.

The method needs the left and right eigenvectors belonging to the $m$
critical eigenvectors,
\begin{align}
	A R_m &= R_m \Lambda_m \: , \qquad
	L_m^\dagger A = \Lambda_m L_m^\dagger \:,
\end{align}
where $\Lambda_m$ is the diagonal eigenvalue matrix for the $m$
critical eigenvalues and $R_m=(r_1,\ldots,r_m)$ and
$L_m=(\ell_1,\ldots,\ell_m)$ are the matrices of right and left
eigenvectors, respectively. These matrices can be made bi-orthonormal,
i.e., $L_m^\dagger R_m = I_m$, such that $R_m L_m^\dagger$ is an
oblique projector on the subspace $\Omega_m$ spanned by the right
eigenvectors.

If the source vector is decomposed as
\begin{align}
b = b_{\parallel} + b_{\ominus} \:,
\end{align}
where $b_{\parallel} = R_m L_m^\dagger b$ is an oblique projection of $b$ on $\Omega_m$ and $b_{\ominus} = b-b_{\parallel}$, then the matrix function can be written as
\begin{align}
f(A) b =  \underbrace{f(A) R_m L_m^\dagger b}_\text{exact}
\:\:\:\: + \underbrace{f(A) b_{\ominus}}_\text{approximation} \:.
\end{align}
The contribution of the first term can be computed exactly, while the
second term can be approximated by the Arnoldi method described in
Sec.~\ref{Sec:Arnoldi} applied to the Krylov subspace
$K_k(A,b_{\ominus})$.  The final approximation is then given by
\begin{align}
f(A) b \approx  R_m f(\Lambda_m) L_m^\dagger b  + |b_{\ominus}| V_k
f(H_k) e_1 \:. 
\label{Arnapp}
\end{align}
As before the function $f(H_k)$ of the internal matrix should be
computed with a suitable method.  The critical eigenvalues with their
left and right eigenvectors should be computed once for all source
vectors $b$.

\section{Two-sided Lanczos approximation}
\label{Sec:2sL}

The Arnoldi method described in Sec.~\ref{Sec:Arnoldi} suffers from
the long recurrences used to orthogonalize the Arnoldi basis. As an
alternative we now consider the two-sided Lanczos method which uses
short recurrences at the cost of giving up orthogonality for
bi-orthogonality.

Consider the two Krylov subspaces $K_k(A,b)$ and $K_k(A^\dagger,b)$
and construct two bi-orthogonal bases $V_k$ and $W_k$ such that
$W_k^\dagger V_k = I_k$ and the matrix $G_k \equiv W_k^\dagger A V_k$
is tridiagonal with
\begin{align}
G_k \equiv W_k^\dagger A V_k = 
\begin{pmatrix}
\alpha_1 & \gamma_1 & 0 & \cdots & 0 \\
\beta_1 & \alpha_2 & \ddots & \ddots & \vdots \\
0 & \ddots & \ddots & \ddots & 0 \\
\vdots & \ddots & \ddots & \ddots & \gamma_{k-1} \\
0 & \cdots & 0 & \beta_{k-1} & \alpha_k \\
\end{pmatrix} .
\end{align}
It can be shown that $V_k$ and $W_k$ can be built using the short
recurrence relations
\begin{align}
\left\{
\begin{aligned}
\beta_i v_{i+1} &= (A - \alpha_i) v_i  - \gamma_{i-1} v_{i-1} \:,\\
\gamma_i^* w_{i+1} &= (A^\dagger - \alpha_i^*) w_i  - \beta_{i-1}^* w_{i-1} \:,
\end{aligned}\right.
\end{align}
where $v_1=w_1=b/|b|$, $\alpha_i = w_i^\dagger A v_i$, and $\beta_i$
and $\gamma_i$ are determined from the normalization condition
$w_{i+1}^\dagger v_{i+1} = 1$.

The matrix $V_k W_k^\dagger$ is an oblique projector on $K_k(A,b)$,
such that an oblique projection of $f(A)b$ on $K_k(A,b)$ is obtained
using
\begin{align}
	y \approx y_\text{obl}=V_k W_k^\dagger f(A) V_k W_k^\dagger b \:.
\end{align}
In analogy to the Arnoldi method we introduce the approximation 
$W_k^\dagger f(A) V_k \approx f(G_k)$
such that
\begin{align} 
y_\text{obl} \approx \tilde y = |b| V_k f(G_k) e_1 \:.
\end{align}
The approximation $\tilde y \in K_k(A,b)$, and the problem is now
reduced to the computation of $f(G_k)$ with $\dim G_k \ll \dim A$.

If the matrix $A$ has small eigenvalues, deflation will again be
necessary when computing the sign function. To implement the
LR-deflation in this case one constructs two bi-orthogonal bases $V_k$
and $W_k$ in $K_k(A,b_\ominus^R)$ and $K_k(A^\dagger,b_\ominus^L)$,
where the directions along $R_m$, respectively $L_m$, have been fully
deflated from $b$, i.e., $b_\ominus^R = (1-R_m L_m^\dagger) b$ and
$b_\ominus^L = (1-L_m R_m^\dagger) b$.  With LR-deflation the function
approximation becomes
\begin{align}
f(A) b \approx  R_m f(\Lambda_m) L_m^\dagger b  + |b_\ominus^R| V_k
f(G_k) e_1 \:. 
\label{2sLapp}
\end{align}

\section{Results}
  
In this section we compare the results obtained with both methods.  In
an initial deflation phase the right and left eigenvectors of
$\gamma_5\Dw(\mu)$ corresponding to the eigenvalues with smallest
magnitude are determined using ARPACK.\footnote{Typical deflation times
  on an Intel Core 2 Duo 2.33GHz workstation were $t=27.5 \text{s}$
  for $m=32$ on the $4^4$ lattice and $t=1713 \text{s}$ for $m=128$ on
  the $6^4$ lattice.}  With these exact eigenvectors the final
approximation to $\Do(\mu) b$ of Eq.~\eqref{Dovmu} is computed as
described in the previous sections.

\begin{figure}
\centerline{\includegraphics[width=0.5\textwidth]{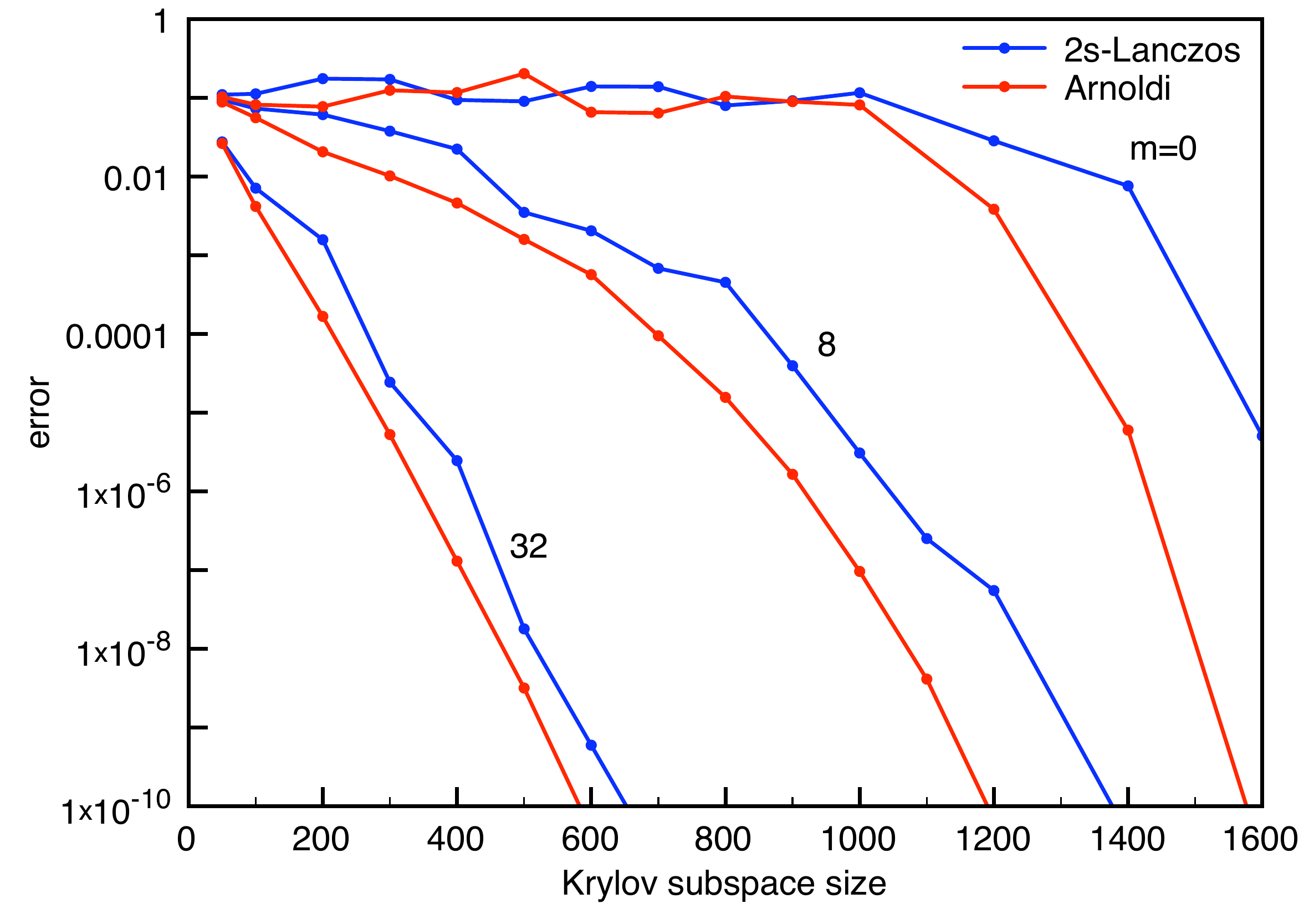}
\includegraphics[width=0.5\textwidth]{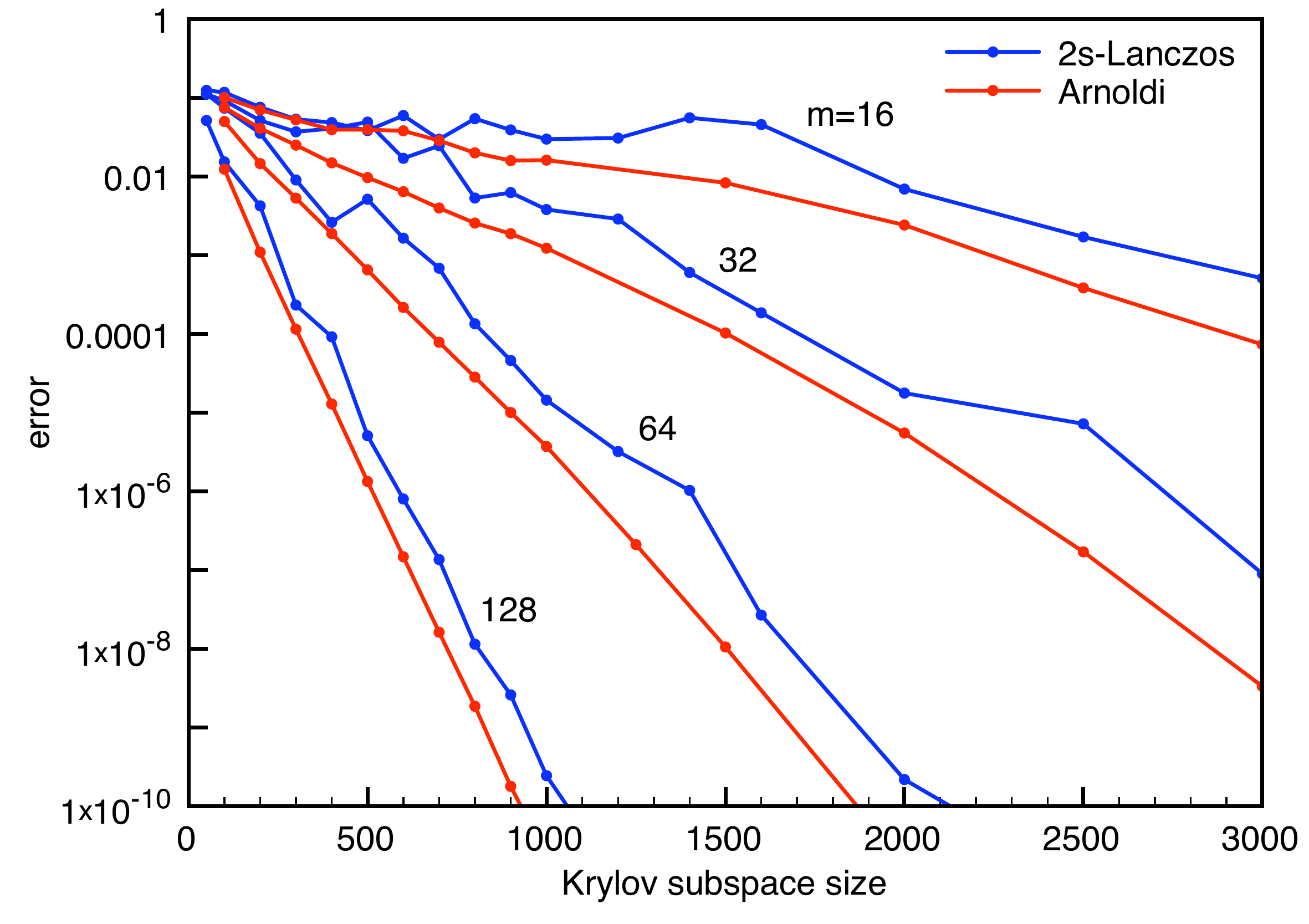}}
\caption{Comparison of the convergence rate for the Arnoldi and the two-sided Lanczos method for a $4^4$ (left) and $6^4$ (right) lattice, for various deflation sizes.}
\label{Fig:cmpconv}
\end{figure}

In Fig.~\ref{Fig:cmpconv} we compare the convergence rate for both methods, by plotting the accuracy versus the Krylov subspace size, and observe that the Arnoldi method has a somewhat better accuracy than the two-sided Lanczos method. To achieve the same accuracy, the Krylov subspace of the two-sided Lanczos method has to be chosen about 20\% larger than in the Arnoldi method. 

\begin{figure}
  \centerline{\includegraphics[width=0.5\textwidth]{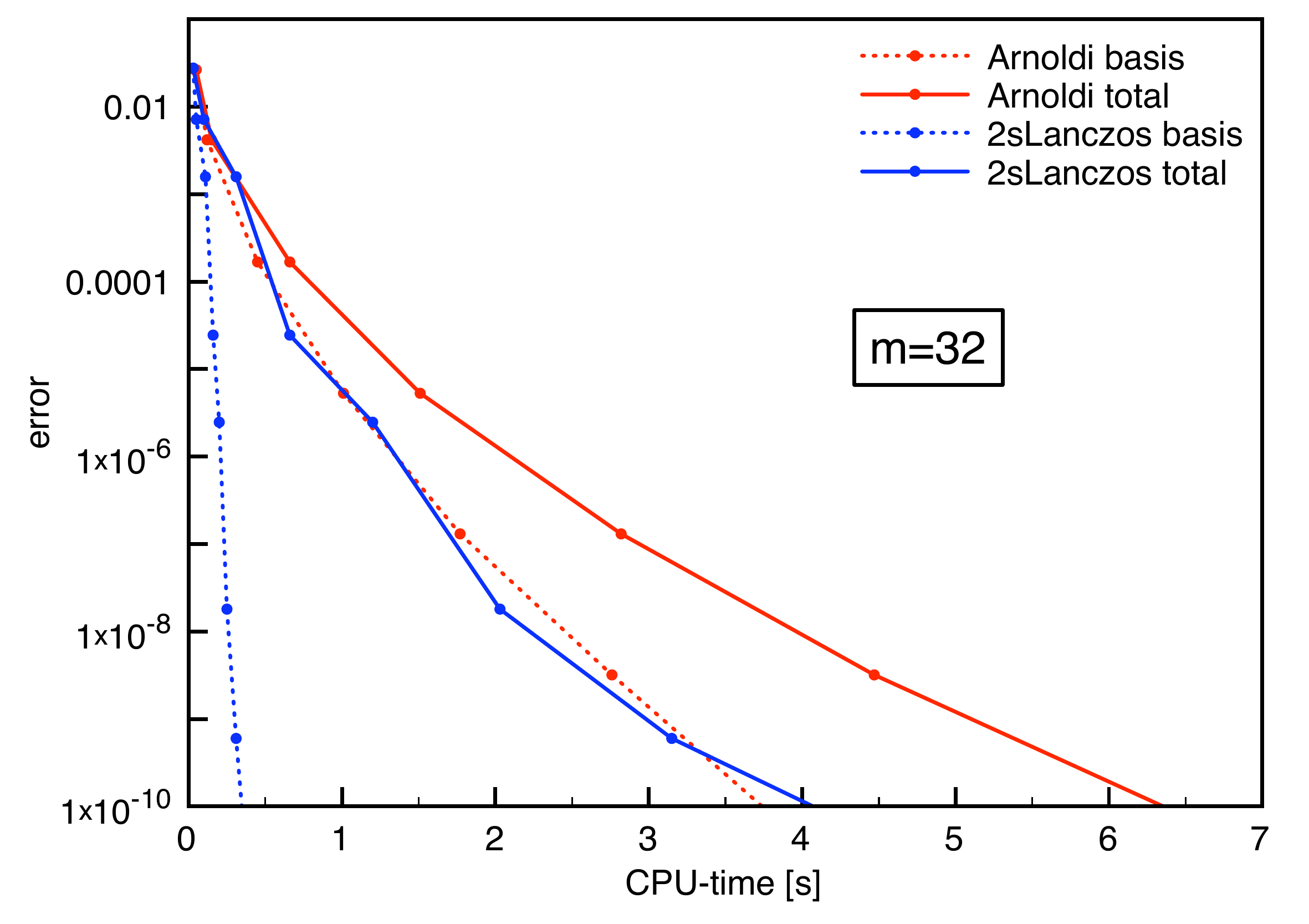}
    \includegraphics[width=0.5\textwidth]{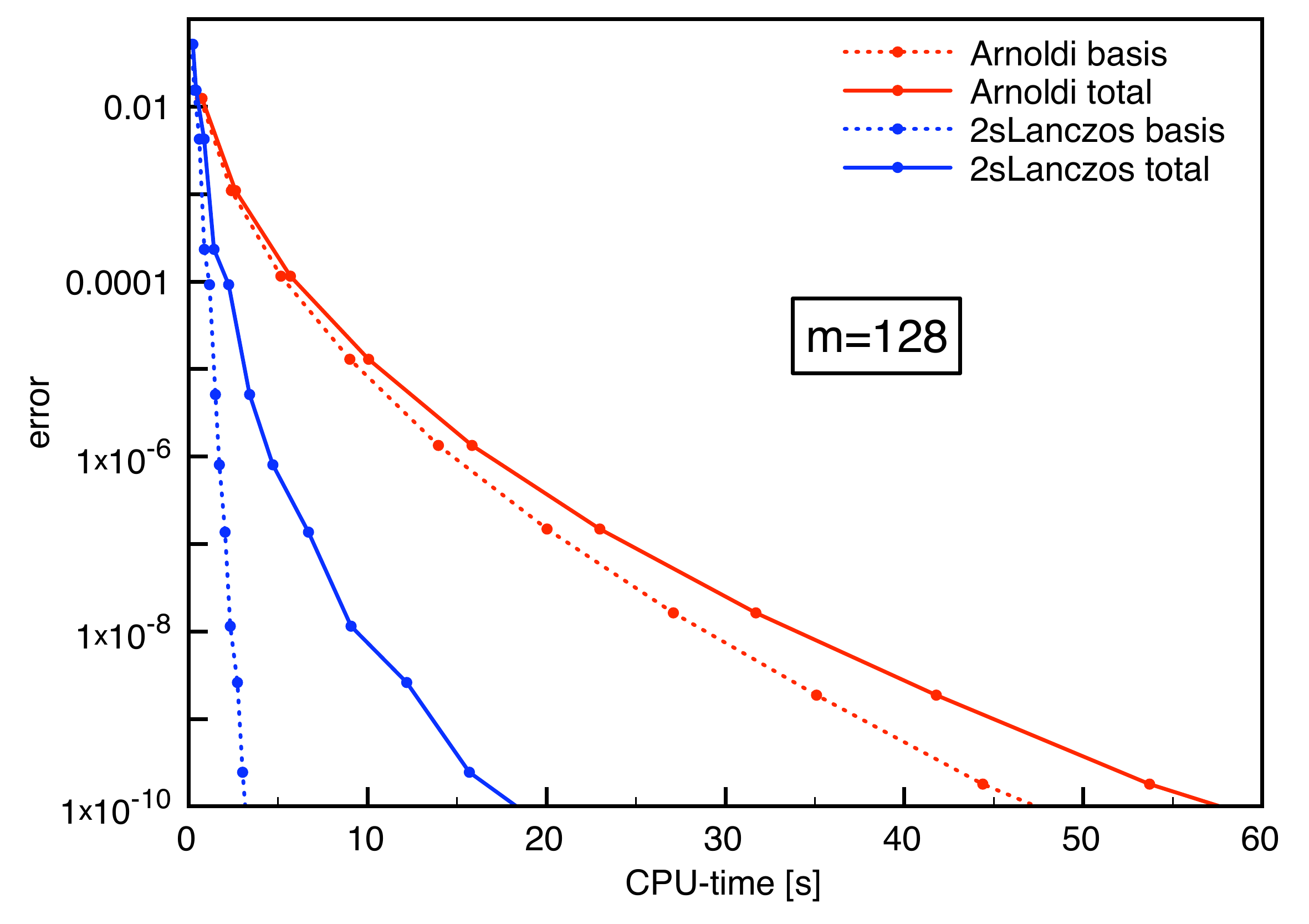}}
  \caption{Accuracy versus CPU-time (in seconds), on an Intel Core 2
    Duo 2.33GHz workstation, for the Arnoldi (red) and the two-sided
    Lanczos method (blue) for a $4^4$ (left) and $6^4$ (right)
    lattice.}
  \label{Fig:cmpeff}
\end{figure}

However, the speed of the short recurrences by far makes up for this
as can be seen from Fig.~\ref{Fig:cmpeff}, where we show the accuracy
as a function of the required CPU-time.  The two-sided Lanczos method
is clearly more efficient than the Arnoldi method, and the advantage
gained by the short recurrences increases as the volume grows.  The
dotted lines show the time needed to build the basis in the Krylov
subspace, while the full lines represent the total CPU-time used by
the iterative method (without the deflation time).  The construction
of the basis is much faster for the two-sided Lanczos method ($\sim N
k$) than in the Arnoldi case ($\sim N k^2$). For the Arnoldi method
the time needed to construct the Arnoldi basis is dominating, while
for the two-sided Lanczos this time can almost be neglected compared
to that needed to compute the sign function of the inner matrix.
Methods to boost the computation of the sign function of the inner
matrix are the subject of a current study.  An additional advantage of
the short recurrences is their possible implementation with small
memory footprint if a two-pass procedure is used to compute
Eq.~\eqref{2sLapp}.

To make the method practical for large-scale lattice simulations it
will be important to improve the deflation phase, even though first
tests on larger lattices ($16^3\times 32$) with realistic parameter
values seem to indicate that the number of deflated eigenvalues can be
taken much smaller than in the $4^4$ and $6^4$ lattices investigated
herein.

The methods discussed above are also currently being tested to compute
eigenvalues of the overlap operator. First results are encouraging and
clearly show the superiority of the two-sided Lanczos method.

\section*{Acknowledgments}
This work was supported in part by DFG grant FOR465-WE2332/4-2.  We
gratefully acknowledge helpful discussions with A. Frommer.

\bibliographystyle{h-elsevier3}

\bibliography{biblio}  

\end{document}